\documentclass[prb,twocolumn,showpacs,preprintnumbers,amsmath,amssymb]{revtex4-1}
\usepackage{graphicx}% Include figure files
\usepackage{dcolumn}% Align table columns on decimal point
\usepackage{bm}% bold math
\usepackage{color}
\usepackage{soul}
\usepackage{amsmath}
\usepackage{multirow}

\begin{document}

 \preprint{APS/123-QED}
\title{ Comment on ``Entropic Costs of Extracting Classical Ticks from a Quantum Clock''}

%\title{Modeling and control epidemic spreading of an SIR model in a residential university environment}

\author{Longyan Gong  $^{1,2}$}
\thanks{Email address: lygong@njupt.edu.cn.}

\affiliation{
 $^{1}$College of Science, Nanjing University of Posts and Telecommunications, Nanjing, 210003, China\\
 $^{2}$Jiangsu Provincial Engineering Research Center of Low Dimensional Physics and New Energy, Nanjing, 210003, China\\
 }
\date{today}
\begin{abstract}
A recent Letter by Wadhia et al. reports a realization of a quantum clock using a double quantum dot (DQD)~\cite{WA25}. This Comment identifies two fundamental issues: (I) the claimed ``quantum clock" exhibits only classical behavior and lacks intrinsic temporal correlations between ticks; (II) the thermodynamic analysis misassigns entropy production and conflates amplification with measurement. The reported combined entropy is an engineering dissipation, not a fundamental cost of quantum timekeeping.
\end{abstract}

%\pacs{71.23.An, 72.15.Rn, 71.30.+h, 71.23.Ft}
%71.23.An Theories and models; localized states
%72.15.Rn Localization effects (Anderson or weak localization)
%71.30.+h Metal-insulator transitions and other electronic transitions
%71.23.Ft Quasicrystals
%-------------------------------
\maketitle
\section*{Part I}

In Ref.~\cite{WA25}, while the experimental data are technically sound and the study is interesting, the claim of ``a quantum clock'' is fundamentally unsubstantiated: the clock exhibits classical behavior; the Letter does not track the time evolution of a single system, thus in principle, the ticks are independent random events lacking intrinsic temporal correlations.

First, the experiment centers on a DQD, augmented by charge sensors and controlled thermal environment.
\begin{table}[!htbp]%[htbp]
\centering
\caption{Three-component decomposition}
\label{tab:decomposition}   % Ìí¼Ó±êÇ©
\renewcommand{\arraystretch}{1.1}  % Ôö¼ÓÐÐ¼ä¾à£¨Ä¬ÈÏ1.0£¬1.5Îª1.5 ±¶£©
\setlength{\tabcolsep}{6pt}        % Ôö¼ÓÁÐ¼ä¾à£¨Ä¬ÈÏ6pt£¬¸ÄÎª12pt£©
%\begin{tabular}{p{2.5cm}p{3.5cm}p{2.8cm}p{4.5cm}}
\begin{tabular}{p{1.5cm}p{4.7cm}p{1.2cm}}
\hline\hline
\textbf{Component} & \textbf{Role} & \textbf{Quantum Nature}\\
\hline
DQD & generates tunneling events & coherence suppressed \\
Charge sensors & convert charge occupation states into classical signals (dc or rf)& No  \\
Thermal environment & source, drain& No \\
\hline\hline
\end{tabular}
\end{table}

 As Table \ref{tab:decomposition} shows, the DQD is quantum hardware that generates tunneling events, but quantum coherence is suppressed under experimental condition; the charge sensors---the direct current (dc) charge sensor and the radio-frequency (rf) reflectometry---are classical devices; the thermal environment is classical. Charge occupation states are $|L\rangle$, $|R\rangle$ and $|0\rangle$, which means a charge in the left dot, in the right dot, and no charge in either dot. The authors explicitly point out that ``superpositions of DQD charge states are not considered as part of this experiment because they decohere faster than the measurement rate. Thus, the DQD is always in a definite charge occupation state when measured.''

The authors acknowledge their demonstration uses ``incoherent semiclassical dynamics.'' Nevertheless, they go on to claim that their conclusions ``also hold for an adaptation to coherent dynamics'' and cite the Supplemental Material (SM). However, the SM contains no derivation, simulation, or experimental demonstration of coherent dynamics, leaving this assertion unsupported.

In short, coherence is suppressed, superpositions are not considered, and no other quantum information resource is utilized, though quantum tunneling generates the ticks. Strictly speaking, this is a clock exhibiting classical behavior, albeit implemented on quantum platforms---physical resource.

Second, a sound definition of a clock is given by Peres~\cite{PE80}: ``A clock is a dynamical system which passes through a succession of states at constant time intervals.'' Foundational approaches like the Page-Wootters mechanism~\cite{PA83} and the Salecker-Wigner prescription~\cite{SA58} define a quantum clock via coherent evolution of a \textit{single} quantum system, where successive states are inherently correlated.

By contrast, the Letter does not track the time evolution of a single system. The DQD is an open system, which exchanges charges
and energy with its thermal environment. Whether holes or electrons are involved, the mobile charge is an electron. Even if the charge occupation state appears identical across cycles, the spin of an incoming electron may differ from that of an outgoing one, meaning the quantum state differs. Thus, the system is not the same across different charge entry-exit cycles. Even with identical spins, this is not continuous evolution of a single quantum state. In principle, no intrinsic temporal correlations exist between transition events from distinct quantum systems.

The Letter uses two estimators: $\Theta_{\text{net}}$ and $\Theta_{\text{opt}}$.  The first one treats a forward (backward) tick as a sequence of transitions from source (drain) to drain (source). Non-tick events are interleaved between ticks. These tick and non-tick events are random. The estimator $\Theta_{\text{net}}[s(t)] = \nu^{-1} N[s(t)]$, where $s(t)$ is the sequence of charge occupation states up to lab time $t$, $N[s(t)]$ is the net number of ticks (one forward tick is counted as $+1$ and one backward tick as $-1$), and $\nu$ is the average charge transfer rate through the DQD. As discussed above, there is no correlation between two successive ticks due to the distinct quantum systems involved. Discarding non-tick events not only weakens potential correlations but also reduces accuracy, as these events also track the progression of time. This DQD device has no fundamental dynamical relation between observed events and physical time. Its time estimation relies entirely on transition rates calibrated against a lab clock, and thus cannot keep time independently.

The second estimator, $\Theta_{\text{opt}}[s(t)] = \sum_{s \in \{0,R,L\}} \frac{n_s(t)}{\Gamma_s}$, counts all state transitions as ticks, where $n_s(t)$ is the number of times the state $s$ appears in the sequence $s(t)$, and $\Gamma_s$ is the escape rate from state $s$. But the core problem remains: transitions within one charge cycle may be correlated, yet ticks across independent entry-exit cycles involve distinct quantum systems and are uncorrelated.

The research group behind the Letter has recently demonstrated that quantum correlated ticks can yield exponential improvement in clock precision~\cite{ME26}. To achieve quantum advantage, this is what quantum clocks should possess and utilize~\cite{PE80,PA83,SA58}. Perhaps as the ticks lack usable temporal correlations, the Letter resorts to the simplest estimators, relying only on mean charge transfer (escape) rates. Indeed, estimating time using the mean rates of random ticks is the most basic, simplest, and crudest approach in statistics. This so-called ``optimal time estimation" rests on the Markov process assumption of exponential waiting-time distributions~\cite {PR25}. It is nothing but a trivial result derived from elaborate mathematical reasoning.

The two estimators will fail if the underlying process is non-stationary or non-Markovian, for instance when the tick interval distribution is too broad or heavy-tailed. The Letter itself shows that non-stationary behavior can occur (Fig. 11 in the SM). In fact, if uncorrelated random ticks were sufficient for accurate time estimation, nearly any time-varying system would qualify as a good clock---an untenable conclusion.\\

\section*{Part II}

In Ref.~\cite{WA25}, Wadhia et al. measure two quantities: $\Sigma_\text{tick} =e|V_\text{DQD}|/k_BT$, which they call the ``entropic cost per tick'' of the DQD clockwork, and the power $P$ dissipated by the charge sensor in dc and rf readout modes. The entropy produced by the amplification and measurement is characterized by $P/T$; for simplicity, we refer to it as combined entropy. Finding that $P/T$ exceeds $\Sigma_\text{tick}$ by nine orders of magnitude, they conclude that ``the entropy produced by the amplification and
measurement of a clock' sticks, which has often been ignored in the literature, is the most important and
fundamental thermodynamic cost of timekeeping at the quantum scale.'' In fact, the combined entropy has been ignored for good reason: it is not a fundamental thermodynamic cost of timekeeping but an engineering dissipation specific to the experimental implementation. The authors' conclusion rests on a misidentification of what constitutes the fundamental entropy cost.

First, misassignment of entropy production of clockwork.  The Letter attributes $\Sigma_{\text{tick}}$ to the DQD clockwork. The DQD clockwork has three charge states $\{|0\rangle, |L\rangle, |R\rangle\}$, and the source and drain act as thermal reservoirs that drive charges into and out of the clockwork. The dissipative steps are the transitions $|0\rangle \rightarrow |L\rangle$ and $|R\rangle \rightarrow |0\rangle$ mediated by the reservoirs; the internal dynamics $|L\rangle \leftrightarrow |R\rangle$ in DQD is coherent tunneling without dissipation. The quantity $\Sigma_{\text{tick}}$ is the Clausius entropy production of these reservoirs per transferred charge. This entropy sustains the clock's operation. The DQD itself is a passive transport channel and undergoes no net entropy accumulation.

Crucially, the Letter itself shows that when $V_{\text{DQD}} = 0$ and $\Sigma_{\text{tick}} = 0$, the DQD still exhibits stochastic jumps among $\{|0\rangle, |L\rangle, |R\rangle\}$, and time can still be estimated via $\Theta_{\text{opt}}$. The clockwork functions identically to the biased case, with zero entropy contribution in both cases. Therefore, $\Sigma_{\text{tick}}$ (experimentally $\sim 60\,k_B$ per tick) is an engineering parameter, not a necessary thermodynamic cost, which can be largely reduced with good experimental designs.

Second, misidentification of the fundamental entropic cost of reading ticks. The Letter combines the ``entropy produced by the amplification and measurement'' as a single quantity to estimate entropic cost of reading ticks. However, these two contributions must be distinguished.

The measurement itself is non-invasive, as capacitive coupling nearly does not change the quantum state of the DQD. Such measurement can in principle be made very small entropic cost~\cite{Bennett1982,Sagawa2009}. Mancino \emph{et al.} experimentally demonstrated that generalized quantum measurements incur a minimal irreversible entropy that can be very small for weakly invasive measurements~\cite{Mancino2018}. The Letter reports combined entropy $P/T$ ($\sim10^9$--$10^{11} k_B$) far exceeds the theoretical lower bound of measurement. Any significant entropy in the combined quantity must therefore come from amplification. But amplification is an engineering choice, not a fundamental entropic cost of generating a record.

Neither $\Sigma_{\text{tick}}$ nor the combined entropy $P/T$ is a fundamental quantity---both are engineering parameters. No general physical principle can be derived from such a comparison, as in practice, engineering dissipation varies significantly with different experimental schemes. The results of Ref.~\cite{WA25} describe a specific experimental implementation, not a fundamental thermodynamic cost of quantum timekeeping. So the conclusion about ``the most important and
fundamental thermodynamic cost'' is unfounded.

%\vspace*{\baselineskip}
%\noindent Longyan Gong$^{1,2}$\\
%\noindent $^{1}$College of Science, Nanjing University of Posts and Telecommunications, Nanjing, 210003, China\\
%\noindent $^{2}$New Energy Technology Engineering of Jiangsu Province, Nanjing University of Posts and %Telecommunications, Nanjing, 210003, China\\
%\noindent Corresponding author: lygong@njupt.edu.cn\\

\end{document}